\documentclass[twocolumn,showpacs,preprintnumbers,amsmath,amssymb,floatfix]{revtex4}

\usepackage{graphicx}
\usepackage{dcolumn}
\usepackage{bm}
\usepackage{amsfonts}

\DeclareMathAlphabet{\mathsfsl}{OT1}{cmr}{bx}{it}
\begin{document}
\title{Rate-dependent slip boundary conditions for simple fluids}
\author{Nikolai~V.~Priezjev}


\affiliation{Department of Mechanical Engineering, Michigan State
University, East Lansing, MI 48824}
\date{\today}
%
\begin{abstract}

The dynamic behavior of the slip length in a fluid flow confined
between atomically smooth surfaces is investigated using molecular
dynamics simulations. At weak wall-fluid interactions, the slip
length increases nonlinearly with the shear rate provided that the
liquid/solid interface forms incommensurable structures. A gradual
transition to the linear rate-dependence is observed upon increasing
the wall-fluid interaction. We found that the slip length can be
well described by a function of a single variable that in turn
depends on the in-plane structure factor, contact density and
temperature of the first fluid layer near the solid wall. Extensive
simulations show that this formula is valid in a wide range of shear
rates and wall-fluid interactions.

\end{abstract}



\maketitle

\section{Introduction}

The interest in modeling of fluid flow in confined geometries has
recently been revived because of the need for optimal design of
micro- and nanofluidic devices~\cite{Darhuber05}. For systems with
large surface to volume ratio, the fluid flow can be significantly
affected by slip at the liquid/solid interface. The existence of
slip and its degree strongly depend on structural and dynamical
properties of the interface. The most commonly used Navier model for
the partial slip boundary conditions states that the liquid slip
velocity is proportional to the rate of shear normal to the surface.
The proportionality coefficient, so-called slip length, is defined
as an extrapolated distance from the wall where the fluid tangential
velocity component vanishes. Alternatively, a ratio of fluid
viscosity to the slip length determines a friction coefficient at
the liquid/solid interface, which relates the interfacial shear
stress and fluid slip velocity. The slip is augmented for specially
designed superhydropobic surfaces~\cite{Rothstein04,Vorobieff06},
high polymer weights~\cite{LegerPRL93,MackayVino}, hydrophobic
surfaces with trapped nanobubbles~\cite{Ishida00,Tyrrell01,Steitz03}
and high shear rates~\cite{Nature97,Granick01,CraigPRL01,Breuer03}.
On the other hand, surface roughness~\cite{Granick02,Leger06} and
hydrophilic surfaces~\cite{Charlaix05,Vinograd06} usually lead to a
reduction of slip. In spite of the long standing interest in slip
behavior, it is not yet clear how the slip length depends on the
local shear rate and on microscopic parameters of the interface.

In the last two decades, molecular dynamics (MD) simulations were
used to examine flow boundary conditions for simple fluids confined
between atomically flat
walls~\cite{Thompson90,KB89,Barrat94,Barrat99,Barrat99fd,Cieplak01,Quirke01}.
The boundary conditions are very sensitive to the wetting properties
and molecular roughness of the surface, as well as to the liquid
structure near the wall. In general, the slip is enhanced for weak
wall-fluid interactions and incommensurable periodic structures of
the surface potential and the first fluid layer. The slip length was
found to correlate with the degree of the surface induced order in
the adjacent fluid layer and wall-fluid interaction
energy~\cite{Thompson90}. Recently, Barrat and
Bocquet~\cite{Barrat94,Barrat99fd} have performed a detailed
analysis based on the Green--Kubo relation for the friction
coefficient at the interface to derive a scaling relation for the
slip length dependence on density, collective diffusion coefficient
and structure factor of the first fluid layer near the wall at
equilibrium. Dynamics of the first layer of liquid molecules near
the wall is closely related to the friction of a monolayer of
adsorbed particles sliding along a solid
substrate~\cite{Smith96,Tomassone97,Ellis04}. Molecular scale
corrugations reduce the effective slip length in cases of periodic
wall roughness~\cite{Barrat94,Priezjev06}, chemically patterned
surfaces~\cite{Priezjev05}, and atomic roughness due to the variable
size of the wall atoms~\cite{Attard04}.

While most of the studies have investigated how a variation of
surface energy and roughness affects boundary conditions, the
dynamic behavior of the slip length with increasing shear rate has
received much less attention. Difficulties in analysis of the
effective slip arise from a combination of different factors, such
as surface roughness, wettability and rate-dependency, which produce
non-equal or even opposite effects on the fluid flow near the
boundary. For example, in non-wetting systems, a reduction of the
slip length due to surface roughness might be compensated by
rate-dependent effects~\cite{Granick02}. Thus, the understanding of
the dynamic behavior of the slip length is important for both the
interpretation of experimental results for flows past rough
surfaces~\cite{Vinograd06,BonaccursoRev05} and modeling fluid flows
in microfluidic channels~\cite{KarniBeskok}.

Molecular dynamics simulations~\cite{Nature97} of simple fluids
undergoing planar shear flow past atomically smooth surfaces have
shown that the slip length increases nonlinearly with the shear
rate, and the usual Navier slip condition is only valid in a limit
of low shear rates. A later study~\cite{Priezjev04} demonstrated
that nonlinear boundary conditions could also describe the flow of
complex fluids which consist of short polymer chains. A transition
from negative to positive slip with varying shear rate in Poiseuille
flows of simple fluids was observed for atomically rough hydrophilic
surfaces, while at smaller wall-fluid interactions, the slip length
increased approximately linearly with shear rate~\cite{Fang05}.
Experimental studies have also reported rate-dependent slip for
Newtonian liquids in pressure driven flows in hydrophobic
microchannels~\cite{Breuer03} and thin film drainage in the surface
force apparatus~\cite{Granick01}. Currently, there is no consensus
regarding the functional form of the rate-dependent slip length and
the existence of a shear rate threshold. As a consequence, this
prevents the analysis of more complex systems involving combined
effects of surface roughness, wettability and rate-dependency.

The focus of this paper is to explore the influence of the
wall-fluid interaction energy and shear rate on slip flow of simple
fluids driven by a constant force. We will show that for strong
wall-fluid interactions the slip length increases linearly with the
shear rate provided the liquid/solid interface forms incommensurable
structures. A gradual transition in rate-dependence of the slip
length, from linear to highly nonlinear, is observed upon reducing
the strength of wall-fluid interactions. A detailed analysis of the
fluid structure near the solid wall shows that in a wide range of
shear rates and wall-fluid interactions the slip length can be
expressed as a function of a single variable that depends on the
in-plane structure factor, contact density and temperature of the
adjacent fluid layer.

This paper is organized as follows. In the next section, we describe
details of molecular dynamics simulations. Predictions from the
continuum hydrodynamics are briefly summarized in
Section~\ref{sec:Hydrodynamics}. Simulation results for the fluid
structure and the slip length are presented in
Section~\ref{sec:Results}. The summary and conclusions are given in
the last section.

\section{Simulation model}
\label{sec:Model}

The computational domain consists of a monoatomic fluid confined
between two atomistic walls. The fluid molecules interact through
the pairwise Lennard-Jones (LJ) potential
\begin{equation}
V_{LJ}(r)\!=4\,\varepsilon\,\Big[\Big(\frac{\sigma}{r}\Big)^{12}\!-\Big(\frac{\sigma}{r}\Big)^{6}\,\Big],
\end{equation}
where $\varepsilon$ and $\sigma$ represent the energy and length
scales of the fluid phase. For computational efficiency the cutoff
distance is set to $r_c\,{=}\,2.5\,\sigma$. The LJ wall-fluid
interaction energy $\varepsilon_{\rm wf}$ and the length scale
$\sigma_{\rm wf}$ are measured in units of $\varepsilon$ and
$\sigma$, respectively. In all our simulations, wall atoms do not
interact with each other and their diameter $\sigma_{\rm w}$ is
equal to $\sigma$. A constant volume accessible to $N\!\,{=}\,3456$
molecules corresponds to the fluid density
$\rho\,{=}\,0.81\,\sigma^{-3}$.

The planar Poiseuille flow was generated by a constant external
force in the $\hat{x}$ direction, which was added to the equation of
motion for each fluid molecule. The heat exchange between the fluid
and an external reservoir was regulated by a Langevin thermostat
with a random force and a damping term with friction coefficient
$\Gamma\,{=}\,1.0\,\tau^{-1}$. This value of the friction
coefficient is small enough not to influence significantly dynamics
of fluid molecules~\cite{Grest86,GrestJCP04}. The damping term was
only applied to the $\hat{y}$ coordinate to avoid a bias in the flow
direction~\cite{Thompson90}. All three components of the equations
of motion for a fluid molecule of mass $m$ are given by
\begin{eqnarray}
\label{Langevin_x}
m\ddot{x}_i & = & -\sum_{i \neq j} \frac{\partial V_{ij}}{\partial x_i} + \text{f}_{\text{x}}\,, \\
\label{Langevin_y}
m\ddot{y}_i + m\Gamma\dot{y}_i & = & -\sum_{i \neq j} \frac{\partial V_{ij}}{\partial y_i} + f_i\,, \\
\label{Langevin_z}
m\ddot{z}_i & = & -\sum_{i \neq j} \frac{\partial V_{ij}}{\partial z_i}\,, %
\end{eqnarray}
where $f_i$ is a randomly distributed force with zero mean and
variance, $\langle
f_i(0)f_j(t)\rangle\,{=}\,2mk_BT\Gamma\delta(t)\delta_{ij}$,
determined from the fluctuation-dissipation relation. The
temperature of the Langevin thermostat is set to
$T\,{=}\,1.1\,\varepsilon/k_B$, where $k_B$ is the Boltzmann
constant. The equations of motion are integrated using the fifth
order gear-predictor algorithm~\cite{Allen87} with a time step
$\triangle t\,{=}\,0.002\,\tau$, where
$\tau\,{=}\,\sqrt{m\sigma^2/\varepsilon}$ is the characteristic LJ
time.

The upper and lower walls of the cell each consisted of 648 atoms
distributed between two (111) planes of the face-centered cubic
(fcc) lattice. A fixed wall density $\rho_w\,{=}\,2.73\,\sigma^{-3}$
corresponds to a nearest-neighbor distance $d\,{=}\,0.8\,\sigma$
between equilibrium positions of wall atoms in the $xy$ plane. The
distance between planes containing wall atoms in a contact with the
fluid was set to a constant value of $24.58\,\sigma$. The dimensions
of the cell in the $xy$ plane were fixed to
$25.03\,\sigma\times7.22\,\sigma$. Periodic boundary conditions were
applied along the $\hat{x}$ and $\hat{y}$ directions.

The steady state Poiseuille flow was induced by a constant force in
the $\hat{x}$ direction, while both the lower and upper walls
remained stationary. Initially, fluid molecules were uniformly
distributed on the centers of the fcc lattice. After an
equilibration period of $100\,\tau$, an external force
$\text{f}_{\text{x}}$ was gradually increased from zero to its final
value corresponding to a steady state flow during next $10^3\,\tau$.
After an additional equilibration of about $10^4\,\tau$, fluid
velocity profiles were averaged within slices of the computational
domain of thickness $\Delta z\,{=}\,0.2\,\sigma$ for a time interval
up to $2\cdot10^5\tau$. Fluid density profiles were computed within
slices of thickness $\Delta z\,{=}\,0.01\,\sigma$ for a time period
$10^5\tau$.

Fluid density and interaction parameters used in this study
correspond to the fluid viscosity
$\mu\,{=}\,(2.2\pm0.2)\,\,\varepsilon\tau\sigma^{-3}$, which was
found to be shear rate independent in a range
$\dot{\gamma}\lesssim0.16\,\tau^{-1}$~\cite{Nature97,Priezjev04}.
The upper estimate of the Reynolds number (based on the maximum
difference in fluid velocities at the center and near walls, fluid
viscosity, and channel width) is ${\sf Re}\!\approx\!10$, indicating
laminar flow conditions.

\begin{figure}[t]
\includegraphics[width=8.5cm,height=6.5cm, angle=0]{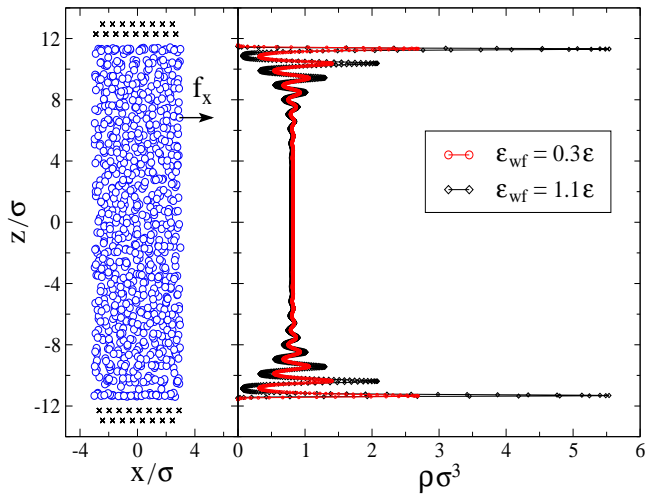}
\caption{(Color online) A snapshot of fluid molecules ($\circ$) and
wall atoms ($\times$) coordinates projected on the $xz$ plane for
$\text{f}_{\text{x}}\,{=}\,0$ and $\varepsilon_{\rm
wf}/\varepsilon\,{=}\,1.1$. Particles positions are only shown in a
range $-3\!\leqslant\!x/\sigma\!\leqslant\!3$ (left). Averaged
density profiles for $\varepsilon_{\rm wf}/\varepsilon\,{=}\,0.3$
($\circ$), $\varepsilon_{\rm wf}/\varepsilon\,{=}\,1.1$ ($\diamond$)
and $\text{f}_{\text{x}}\,{=}\,0.001\,\varepsilon/\sigma$ (right).}
\label{mol_dens}
\end{figure}

\section{Hydrodynamic predictions}
\label{sec:Hydrodynamics}

For the planar Poiseuille flow under an externally applied force
$\text{f}_{\text{x}}$ in the $\hat{x}$ direction, the solution of
the Navier-Stokes equation is described by a parabolic velocity
profile~\cite{KarniBeskok}
\begin{equation}
\textrm{v}(z)=\frac{\rho\,\text{f}_{\text{x}}}{2\mu}\,(h^2-z^2)+V_s\,,
\label{velo_hydro}
\end{equation}
where the fluid viscosity $\mu$ is assumed to be shear rate
independent. The boundary conditions for the fluid velocity are
prescribed at the confining parallel walls,
$\textrm{v}(-h)\,{=}\,\textrm{v}(h)\,{=}\,V_s$. The shear rate at
the liquid/solid interface relates the fluid slip velocity and the
slip length as follows
\begin{equation}
\frac{\partial \textrm{v}}{\partial z}\,(-h)=\frac{V_s}{L_s}\,.
\end{equation}
A quantity of interest for experimental measurements, the flow rate,
can be evaluated by integrating the fluid velocity profile,
Eq.\,(\ref{velo_hydro}), across the channel width
\begin{equation}
Q_{slip}=\int_{-h}^{h}\!{\textrm{v}(z)\,dz}=\frac{2}{3}\,
\frac{\rho\,\text{f}_{\text{x}}h^3}{\mu}+2\,hV_s\,,
\end{equation}
where the second term represents a correction to the flow rate due
to slip boundary conditions. A relative increase in the flow rate
due to slip can also be expressed in terms of the slip length and
the distance between confining walls
\begin{equation}
\frac{Q_{slip}}{Q_{no-slip}}=1+6\,\frac{L_s}{2h}\,.
\label{flow_rate}
\end{equation}
Fluid velocity profiles obtained from MD simulations will be
compared with the hydrodynamic predictions, Eq.\,(\ref{velo_hydro}),
in the next section. Parameters of the liquid/solid interface
correspond to a flow regime, where the slip length is comparable
with the channel width, $2h$; and, therefore, the flow rate strongly
depends on the boundary conditions.

\section{Results}
\label{sec:Results}

\subsection{Fluid structure near the walls}

Dynamical and structural properties of a fluid can be significantly
affected by the presence of a solid substrate~\cite{Israel92}. A
flat solid wall constrains the motion of fluid molecules in a normal
direction and induces oscillations in the fluid density profile.
Typically, these density oscillations gradually decay within a few
molecular diameters away from the wall. The spatial distribution of
molecules near the wall becomes non-isotropic and consists of fluid
layers with thickness of about a molecular diameter. Although a
large amplitude of density oscillations near the wall is a signature
of a high surface attraction energy, the enhanced layering does not
necessarily correlate with a reduction of the fluid slippage at the
interface. For example, a highly attractive surface potential free
of lateral corrugations would induce substantial fluid layering,
which can be interpreted as an infinite slip. In our simulations,
density oscillations relax to a uniform bulk profile within
$5\!-\!6\,\sigma$ away from the wall, see Fig.\,\ref{mol_dens}. As
expected, the higher surface attraction energy causes a more
pronounced fluid layering.

\begin{figure}[t]
\includegraphics[width=8.5cm,height=6.5cm, angle=0]{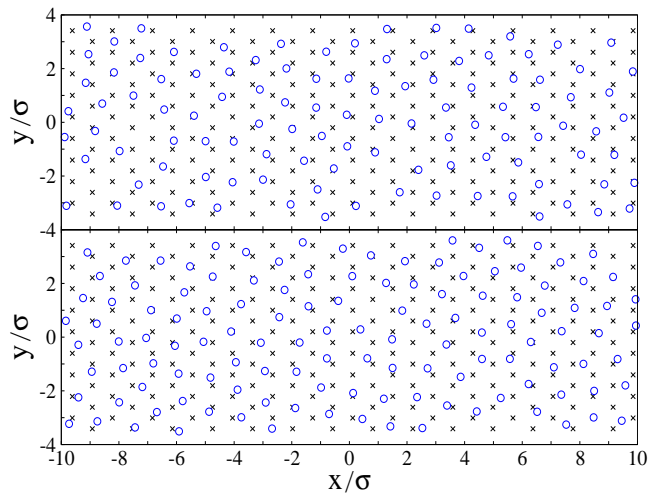}
\caption{(Color online) Instantaneous $x$ and $y$ coordinates of the
fluid molecules ($\circ$) in a contact with the lower wall atoms
($\times$) after an equilibration period of $10^4\tau$  for
$\text{f}_{\text{x}}\,{=}\,0$. Wall-fluid interaction energy is
fixed to $\varepsilon_{\rm wf}/\varepsilon\,{=}\,0.3$ (top) and
$\varepsilon_{\rm wf}/\varepsilon\,{=}\,1.1$ (bottom). The fcc wall
layer is located at $z=-\,12.29\,\sigma$.} \label{layers}
\end{figure}

In general, the fluid layer closest to the flat wall has the largest
degree of in-plane order characterized by the structure factor
$S(\mathbf{k})\,{=}\,1/N_{\ell}\,\,|\sum_j
e^{i\,\mathbf{k}\cdot\mathbf{r}_j}|^2$, where
$\mathbf{r}_j\!=(x_j,y_j)$ is a two-dimensional vector of a molecule
position and $N_{\ell}$ is the total number of molecules within the
adjacent layer. Factors affecting the in-plane order include a
correlation between fluid molecules near the wall and the energy
landscape of the surface potential. The degree of the surface
induced order depends on a mismatch between the wall lattice
constant and the nearest-neighbor distance in the adjacent fluid
layer. If the wall-fluid interaction $\varepsilon_{\rm wf}$ is
comparable with the fluid-fluid energy scale $\varepsilon$, then
commensurable wall-fluid structures would typically result in stick
boundary conditions, or even fluid epitaxial locking, while
incommensurable structures would likely produce more
slippage~\cite{Thompson90,Attard04}.

\begin{figure}[t]
\includegraphics[width=8.5cm,height=6.5cm, angle=0]{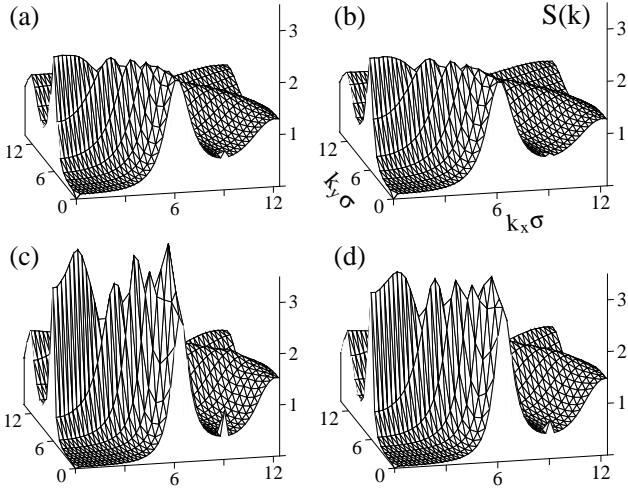}
\caption{Structure factor $S(k_{\text{x}},k_{\text{y}})$ in the
first fluid layer for $\varepsilon_{\rm wf}/\varepsilon\,{=}\,0.3$
(top) and $\varepsilon_{\rm wf}/\varepsilon\,{=}\,1.1$ (bottom). A
force per fluid molecule is
$\text{f}_{\text{x}}\,{=}\,0.001\,\varepsilon/\sigma$ (a),
$0.012\,\varepsilon/\sigma$ (b), $0.001\,\varepsilon/\sigma$ (c),
and $0.025\,\varepsilon/\sigma$ (d). A small peak appears at the
first reciprocal lattice vector
$\mathbf{G}_1\,{=}\,(9.04\,\sigma^{-1},0).$} \label{sk}
\end{figure}

In this study, the liquid and solid phases form an interface with a
mismatch between neighboring distance within the first fluid layer
(of about $\sigma$) and a lattice constant in the $xy$ plane,
$d\,{=}\,0.8\,\sigma$. Figure\,\ref{layers} shows instantaneous
snapshots of molecular positions in a fluid layer adjacent to the
lower wall for the highest ($\varepsilon_{\rm
wf}/\varepsilon\,{=}\,1.1$) and the lowest ($\varepsilon_{\rm
wf}/\varepsilon\,{=}\,0.3$) wall-fluid interaction energies. In the
later case, the averaged structure factor within the first layer
exhibits typical short-range fluid ordering characterized by a
circular ridge at $|\mathbf{k}|\!\approx\!2\pi\!/\sigma$ with an
amplitude $S_1\!\approx\!2.2$, see Fig.\,\ref{sk}. For
$\varepsilon_{\rm wf}/\varepsilon\,{=}\,1.1$, the surface induces
higher short-range order, which is enhanced along the crystal axes
in the $xy$ plane. The height of the largest peak in
Fig.\,\ref{sk}\,(a) is $S_1\!\approx\!4.1$. A smaller peak of the
in-plane structure factor due to periodic surface potential
corresponds to the first reciprocal lattice vector
$\mathbf{G}_1\,{=}\,(9.04\,\sigma^{-1},0)$, see
Fig.\,\ref{sk}\,(a)\,--\,(d). The amplitude of the peak at
$\mathbf{G}_1$ decreases at larger values of the external force. A
correlation between surface induced order in the first fluid layer
and the slip length will be discussed in the next subsection.

\subsection{Fluid velocity profiles and slip length}

The magnitude of the external force, which is required to reach a
parabolic velocity profile described by Eq.\,(\ref{velo_hydro}),
depends on the fluid viscosity, density and wall-fluid interaction
parameters~\cite{Barrat99,KB89,Cieplak01,Quirke01,Fang05}. Since the
slip velocity is not known {\it a priori}, the value of the force in
MD simulations is usually adjusted so that a fluid velocity profile
could be accurately resolved without an excessive computational
effort due to thermal averaging. One of the goals of this study is
to systematically explore the effect of an applied force on a flow
of simple fluids near solid boundaries and to determine a variation
of the slip length as a function of shear rate. In our simulations,
the channel width, $2h\,{=}\,23.58\,\sigma$, is large enough to
avoid extreme confinement
conditions~\cite{Bitsanis90,Travis97,Travis00}, where deviations
from macroscopic hydrodynamics are expected.

\begin{figure}[t]
\includegraphics[width=8.5cm,height=6.5cm, angle=0]{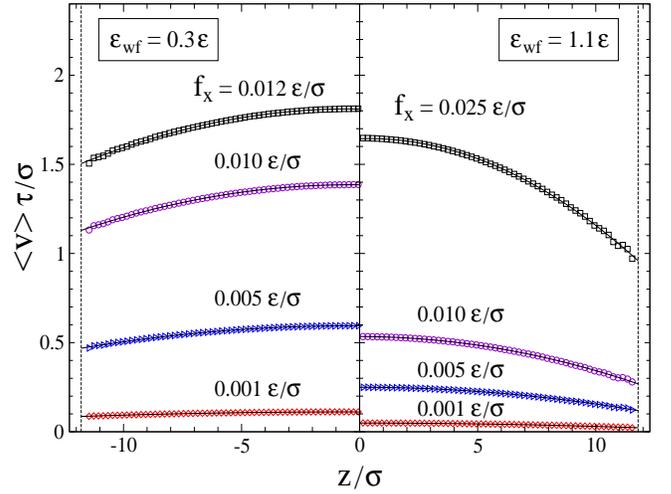}
\caption{(Color online) Averaged velocity profiles,
$\langle\textrm{v}\rangle\,\tau/\sigma$, for indicated values of an
applied force per fluid molecule. Wall-fluid interaction energy is
set to $\varepsilon_{\rm wf}/\varepsilon\,{=}\,0.3$ (left) and
$\varepsilon_{\rm wf}/\varepsilon\,{=}\,1.1$ (right). Solid lines
represent parabolic fit to the data. Dashed lines denote positions
of liquid/solid interfaces. Vertical axes coincide with fcc lattice
planes at $z\,{=}\,\pm 12.29\,\sigma$.} \label{parab_velo}
\end{figure}

Examples of averaged velocity profiles for different values of the
external force $\text{f}_{\text{x}}$ and fixed wall-fluid
interaction energies, $\varepsilon_{\rm wf}/\varepsilon\,{=}\,0.3$
and $\varepsilon_{\rm wf}/\varepsilon\,{=}\,1.1$, are shown in
Fig.\,\ref{parab_velo}. The data are presented only in half of the
channel because of the symmetry with respect to $z\,{=}\,0$ plane,
$\textrm{v}(z)\!=\textrm{v}(-z)$. Fluid velocity profiles are well
fitted by a parabola as expected from the hydrodynamic predictions,
see Eq.\,(\ref{velo_hydro}). Weak oscillations within $2\,\sigma$
near the walls correspond to a pronounced fluid layering
perpendicular to the surface. In a range of wall-fluid interaction
energies considered in this study, $0.3\!\leqslant\!\varepsilon_{\rm
wf}/\,\varepsilon\!\leqslant\!1.1$, fluid flow undergoes slippage at
the solid walls. Fluid velocity in the channel and at the interfaces
increases with the applied force. The shear viscosity,
$\mu\!=\!(2.2\pm0.2)\,\,\varepsilon\tau\sigma^{-3}$, which was
computed from the Kirkwood relation~\cite{Bird87}, remained
independent of the applied force.

A ratio between fluid slip velocity and the local shear rate at the
interface defines the slip length, denoted by $L_s$ throughout. For
the parabolic profiles, the slip length was evaluated by linear
extrapolation of the slope at the interface to zero velocity. In our
simulations, the position of the interface in the $\hat{z}$
direction is defined at a distance $0.5\,\sigma_{\rm w}$ away from
the fcc lattice planes, see vertical dashed lines in
Fig.\,\ref{parab_velo}. This offset was chosen to account for the
excluded volume due to wall atoms. The slip length was defined as
the average of values extracted at the top and the bottom walls.

\begin{figure}[t]
\includegraphics[width=8.5cm,height=6.5cm, angle=0]{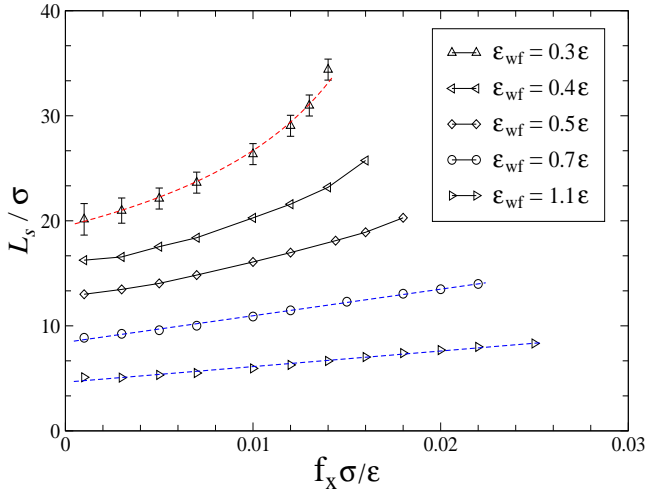}
\caption{(Color online) Variation of the slip length, $L_s/\sigma$,
as a function of an applied force per fluid molecule. Wall-fluid
interaction energy is set to $\varepsilon_{\rm
wf}/\varepsilon\,{=}\,0.3$ ($\vartriangle$), 0.4 ($\triangleleft$),
0.5 ($\diamond$), 0.7 ($\circ$), and 1.1 ($\triangleright$),
respectively. Dashed curve is the best fit to
$L_s(\text{f}_{\text{x}})=L_s^{\circ}~(1-\text{f}_{\text{x}}/\text{f}_{c})^{-0.5}$
with $L_s^{\circ}\!=\!19.5\,\sigma$ and
$\text{f}_{c}\,{=}\,0.021\,\varepsilon/\sigma$. Straight dashed
lines represent the best fit to the data.} \label{f_ls_eps}
\end{figure}

The dynamic response of the slip length as a function of the
external force is presented in Fig.\,\ref{f_ls_eps}. A gradual
transition in the functional dependence of
$L_s(\text{f}_{\text{x}})$ is observed by varying the strength of
the wall-fluid interaction. The slip length increases monotonically
with the applied force for $\varepsilon_{\rm
wf}/\varepsilon\!\geqslant\!0.7$. The data can be well fitted by a
straight line, see Fig.\,\ref{f_ls_eps}. At lower surface energies,
$\varepsilon_{\rm wf}/\varepsilon\!\leqslant\!0.5$, the relation
between $L_s$ and $\text{f}_{\text{x}}$ becomes nonlinear and
exhibits a pronounced downward curvature for $\varepsilon_{\rm
wf}/\varepsilon\,{=}\,0.3$. For each curve shown in
Fig.\,\ref{f_ls_eps}, a ratio of maximum slip length to its value at
small applied forces is equal to $1.63\pm0.13$. This factor
determines an upper bound for the increase in the flow rate due to
slip dependence on the applied force. Given the fixed channel width
used in our study, the maximum relative gain in the flow rate due to
variation of the slip length with the force is $1.59\pm0.08$ for
$\varepsilon_{\rm wf}/\varepsilon\,{=}\,0.3$ and $1.36\pm0.08$ for
$\varepsilon_{\rm wf}/\varepsilon\,{=}\,1.1$. These results suggest
that a significant drag reduction for laminar flows can be achieved
through the increase in the pressure difference across a
microfluidic channel.

\begin{figure}[t]
\includegraphics[width=8.5cm,height=6.5cm, angle=0]{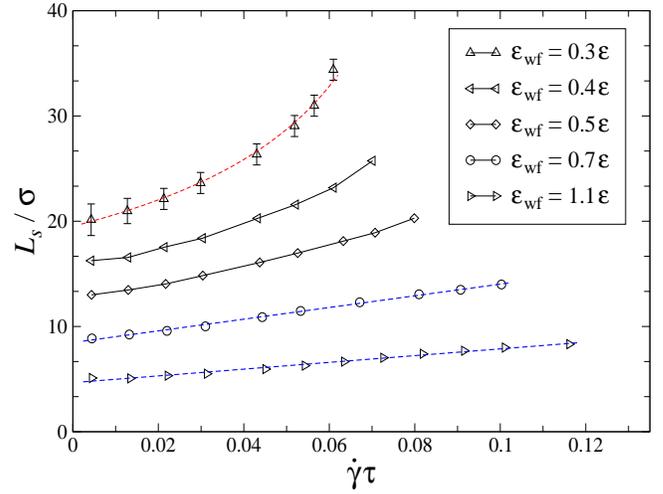}
\caption{(Color online) Behavior of the slip length as a function of
the local shear rate at the interface. Values of the wall-fluid
interaction energy are listed in the inset. The same data as in
Fig.\,\ref{f_ls_eps}. Dashed curve is the best fit to
Eq.\,(\ref{nature_law}) with $L_s^{\circ}\!=\!19.5\,\sigma$ and
$\dot{\gamma}_c\!=0.093\,\tau^{-1}$. Solid curves are a guide for
the eye. Dashed lines show the best linear fit to the data.}
\label{shear_ls_eps}
\end{figure}

In the original paper by Thompson and Troian~\cite{Nature97} on
shear flow of simple fluids, the slip length was found to increase
nonlinearly with the shear rate. In a range of accessible shear
rates and weak wall-fluid interactions, the MD data were well
described by a power law function
\begin{equation}
L_s(\dot{\gamma})=L_s^{\circ}~(1-{\dot{\gamma}}/{\dot{\gamma}_c})^{-0.5},
\label{nature_law}
\end{equation}
where $L^{\circ}_s$ and $\dot{\gamma}_c$ are fitting parameters. In
our simulations, the shear rate is proportional to the external
force, see Eq.\,(\ref{velo_hydro}), and, therefore, the analogous
expression for the slip length dependence on the applied force
should be
$L_s(\text{f}_{\text{x}})=L_s^{\circ}~(1-\text{f}_{\text{x}}/\,\text{f}_c)^{-0.5}$.
This form was used to fit the data for the lowest wall-fluid
interaction energy $\varepsilon_{\rm wf}/\varepsilon\,{=}\,0.3$, see
dashed curve in Fig.\,\ref{f_ls_eps}. The agreement is rather good
in the range of applied forces
$\text{f}_{\text{x}}/\,\text{f}_{c}\!\lesssim0.67$.

For each curve presented in Fig.\,\ref{f_ls_eps}, the external force
was varied from
$\text{f}_{\text{x}}\,{=}\,0.001\,\varepsilon/\sigma$ up to a
maximum value, which depends on $\varepsilon_{\rm wf}$. This value
of the force corresponds to a maximum shear stress the liquid/solid
interface can support. With a further increase of the force, the
fluid flow acquires a large velocity component in the $\hat{x}$
direction, $\langle\textrm{v}\rangle\!\gg v_T$, where
$v^2_{T}\!=k_BT\!/m$ is the thermal fluid velocity. In this extreme
regime, the dynamics of fluid molecules near walls cannot be
resolved accurately with the integration time step used in this
study. We note, however, that test runs with a smaller time step
$\triangle t\!=\!0.001\,\tau$ did not produce noticeable changes in
the results presented in Fig.\,\ref{f_ls_eps}. The transition to the
flow regime characterized by very large slip velocities is not the
main focus of this paper and, therefore, it was not studied in
detail.

\begin{figure}[t]
\includegraphics[width=8.5cm,height=6.5cm, angle=0]{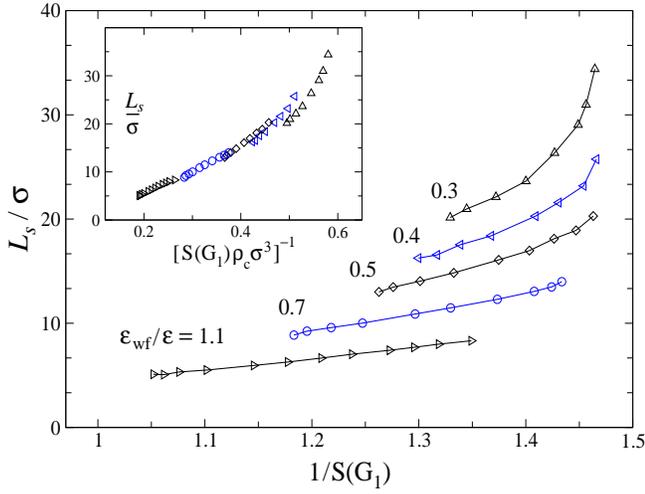}
\caption{(Color online) A correlation between the slip length,
$L_s/\sigma$, and the inverse value of the in-plane structure
factor, $1/S(\mathbf{G}_1)$, evaluated at the first reciprocal
lattice vector. Solid curves are a guide for the eye. The inset
shows the same data plotted as a function of
$[S(\mathbf{G}_1)\,\rho_c\,\sigma^3]^{-1}$.} \label{ls_S9_ro}
\end{figure}

The parabolic shape of the fluid velocity profiles implies that the
external force $\text{f}_{\text{x}}$ is proportional to the
interfacial shear rate, see Eq.\,(\ref{velo_hydro}). The functional
dependence of the slip length, therefore, is expected to be similar
for $\text{f}_{\text{x}}$ and the local shear rate.
Figure\,\ref{shear_ls_eps} shows the same MD data as in
Fig.\,\ref{f_ls_eps}, but replotted in the axis $L_s$ versus local
shear rate, as extracted from parabolic velocity profiles at the
location of interfaces. The range of shear rates is below the values
reported for laminar flows in Ref.\,\,\cite{Thompson90}. The slip
length increases with shear rate, and the growth of $L_s$ is
enhanced at lower values of $\varepsilon_{\rm wf}$. The dependence
$L_s(\dot{\gamma})$ for $\varepsilon_{\rm wf}/\varepsilon\,{=}\,0.3$
can be well fitted by the power law function,
Eq.\,(\ref{nature_law}), with $L_s^{\circ}\,{=}\,19.5\,\sigma$ and
$\dot{\gamma}_c\,{=}\,0.093\,\tau^{-1}$, see dashed curve in
Fig.\,\ref{shear_ls_eps}. Thus, for a weak wall-fluid interaction
our results are in agreement with those reported in a previous
study~\cite{Nature97} on dynamical behavior of the slip length in
boundary driven flows. Furthermore, at higher surface energies,
$\varepsilon_{\rm wf}/\varepsilon\!\geqslant\!0.7$, the slip length
increases linearly with the interfacial shear rate, see
Fig.\,\ref{shear_ls_eps}. In this regime, the simulations results
might be relevant to a monotonic growth of the slip length with the
shear rate measured in pressure driven flows in hydrophobic
microchannels~\cite{Breuer03}. Finally, in contrast to our results,
a transition from the linear to the power law rate-dependence upon
{\it increasing} the strength of wall-fluid interactions was
reported in Ref.\,\,\cite{Fang05}. The difference in the slip
behavior might be explained by the lower fluid density and higher
shear rates examined in that study~\cite{Fang05}.

Molecular-scale corrugations of a solid wall composed of
periodically arranged LJ atoms induce in-plane order in the adjacent
fluid layer. The amount of surface induced order in the first fluid
layer is reflected in the fluid in-plane structure factor. A
correlation between the slip length in a shear rate independent
regime and the peak value of the fluid structure factor evaluated at
the first reciprocal lattice vector was established
earlier~\cite{Thompson90,Barrat99fd}. In this study, the peak of the
structure factor at the first reciprocal lattice vector
$\mathbf{G}_1\,{=}\,(9.04\,\sigma^{-1},0)$ is displaced from the
circular ridge at the vector $|\mathbf{k}|\!\approx\!2\pi\!/\sigma$,
see Fig.\,\ref{sk}. The magnitude of the peak at $\mathbf{G}_1$
decreases with increasing slip velocity. Figure~\ref{ls_S9_ro} shows
the behavior of the slip length as a function of the inverse value
of the structure factor. For the largest wall-fluid interaction
energy $\varepsilon_{\rm wf}/\varepsilon\,{=}\,1.1$ the slip length
increases linearly with $1/S(\mathbf{G}_1)$. At lower surface
energies, the function deviates from the linear dependence and
increases more rapidly for $\varepsilon_{\rm
wf}/\varepsilon\,{=}\,0.3$. These results demonstrate that the slip
length in a \textit{shear rate dependent} regime strongly correlates
with the surface induced order in the fluid layer adjacent to the
wall.

\begin{figure}[t]
\includegraphics[width=8.5cm,height=6.5cm, angle=0]{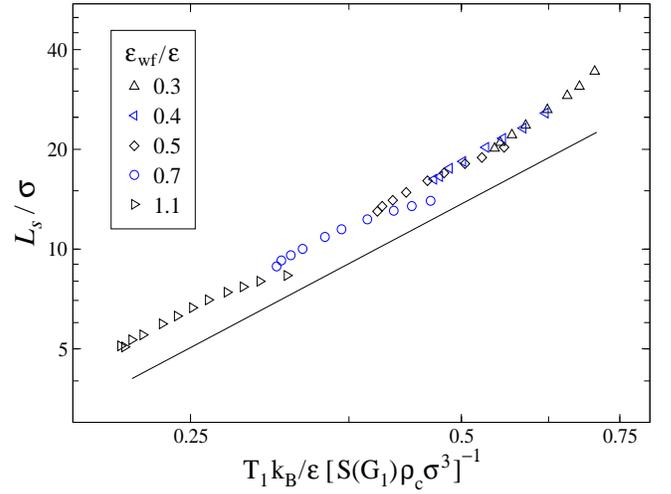}
\caption{(Color online) Log-log plot of the slip length as a
function of a combined ratio
$T_1k_B/\varepsilon\,[S(\mathbf{G}_1)\,\rho_c\,\sigma^3]^{-1}$.
Values of the wall-fluid interaction energy are tabulated in the
inset. The same data as in Fig.\,\ref{f_ls_eps}.
The solid line with a slope $1.44$ is plotted for reference.}
\label{ls_T_S9_ro}
\end{figure}

In a zero shear limit, the Green--Kubo analysis for the friction
coefficient at the liquid/solid interface shows that for attractive
wall-fluid interactions the slip length depends on the structure
factor, contact density, diffusion coefficient and temperature of
the first fluid layer~\cite{Barrat99fd}. In our simulations, the
equilibrium fluid density and temperature profiles near the wall are
modified at higher shear rates. The contact density, $\rho_c$, was
determined from the maximum of the fluid density profile in the
first fluid layer, see Fig.\,\ref{mol_dens}. The contact density
increases with the strength of wall-fluid interaction and decreases
with shear rate. In the inset of Fig.\,\ref{ls_S9_ro} the slip
length is plotted against an inverse product of the structure factor
and the contact density. Except for the lowest wall-fluid
interaction energy, $\varepsilon_{\rm wf}/\varepsilon\,{=}\,0.3$,
the functional form of the slip length consists of nearly linear
interconnected segments each characterized by its own value of
$\varepsilon_{\rm wf}$.

The data for the slip length for different wall-fluid interaction
energies and shear rates can be collapsed onto a single master curve
by taking into account a variation in temperature of the first fluid
layer. In a steady state flow induced by a constant force, fluid
temperature was computed from the kinetic energy
\begin{equation}
k_BT=\frac{m}{3N}\sum_{i=1}^N\,[\dot{\mathbf{r}}_i-\mathbf{v}(\mathbf{r}_i)]^2,
\label{temp3d}
\end{equation}
where $\mathbf{r}_i$ is a three-dimensional vector of a molecule
position and $\mathbf{v}(\mathbf{r}_i)$ is the local average flow
velocity. At low shear rates, $\dot{\gamma}\lesssim0.02\,\tau^{-1}$,
the fluid temperature remains equal to that imposed by the Langevin
thermostat, $T\,{=}\,1.1\,\varepsilon/k_B$, and it increases
slightly at higher $\dot{\gamma}$. The maximum relative increase of
$T$ is about $3.5\%$ for each value of $\varepsilon_{\rm wf}$. The
heatup is larger near the walls because of the higher shear rates
and the slip velocity, which becomes comparable to the thermal fluid
velocity, see Fig.\,\ref{parab_velo}. Temperature of the first fluid
layer, $T_1$, rises by about $10\%$ at the highest shear rates
reported in Fig.\,\ref{shear_ls_eps}.

Figure~\ref{ls_T_S9_ro} shows the slip length as a function of a
combined ratio of temperature to the product $S(\mathbf{G}_1)
\rho_c$ evaluated in the first fluid layer. In a wide range of shear
rates and for $0.3\!\leqslant\!\varepsilon_{\rm
wf}/\varepsilon\!\leqslant\!1.1$, the slip length is well described
by a power law function
\begin{equation}
L_s~\sim~(\,T_1/S(\mathbf{G}_1)\,\rho_c)^{\alpha}, \label{ls_GroT}
\end{equation}
with $\alpha\,{=}\,1.44\pm0.10$. A solid line in
Fig.\,\ref{ls_T_S9_ro} corresponds to the value $\alpha\,{=}\,1.44$.
This result implies that the condition
$T_1/S(\mathbf{G}_1)\rho_c=const$ defines a curve on the plane
$\varepsilon_{\rm wf}$ and $\dot{\gamma}$ characterized by a
constant slip length. The functional dependence of the slip length,
Eq.\,(\ref{ls_GroT}), can also be determined from equilibrium
measurements of $(S(\mathbf{G}_1)\,\rho_c)^{-\alpha}$ for different
surface energies $\varepsilon_{\rm wf}$. This prediction is in
qualitative agreement with previous MD
results~\cite{Thompson90,Barrat99fd}, which demonstrated that the
slip length decreases with the contact density and surface induced
order in the adjacent fluid layer. A strong correlation between the
slip length and microscopic properties of the first fluid layer
provides a framework for the analysis of systems with combined
effects of wettability and rate-dependency.

\section{Conclusions}
\label{sec:Conclusions}

In this paper the effect of surface energy and shear rate on the
slip length in a flow of simple fluids was studied by molecular
dynamics simulations. Fluid velocity profiles in steady state flow
induced by a constant force were fitted by a parabola with a shift
by the value of the slip velocity. The slope of the parabolic fit at
the interface was used to define the local shear rate. For a weak
wall-fluid interaction the slip length increases nonlinearly with
the shear rate and its dependence can be well fitted by a power law
function. Increasing the strength of wall-fluid interactions leads
to the linear rate-dependence of the slip length. For a fixed
channel width, the flow rate increases significantly due to the
rate-dependence of the slip length for both weak and strong
wall-fluid interactions. Simulation results also indicate a strong
correlation between the slip length and the surface induced order in
the first fluid layer in a shear rate-dependent regime. We showed
that in a wide range of wall-fluid interaction energies and shear
rates the slip length is well described by a function of a single
variable that depends on the in-plane structure factor, contact
density and temperature of the first fluid layer.

Future work will show how sensitive these results are to the
variation of molecular-scale roughness. The surface induced order in
the adjacent fluid layer and the slip length might be affected by
the presence of substrate inhomogeneities. The effect of thermal,
random and periodically corrugated surfaces on slip behavior in a
rate-dependent regime should be explored.

\section*{Acknowledgments}
Financial support from the Michigan State University Intramural
Research Grants Program is gratefully acknowledged. The author
thanks S.~M. Troian, A.~A. Darhuber and L. Bocquet for useful
discussions and P.~A. Thompson for kindly sharing his source code.
Computational work in support of this research was performed at
Michigan State University's High Performance Computing Facility.

\bibliographystyle{prsty}

\end{document}